# SCHEDULING AND ALLOCATION ALGORITHM FOR AN ELLIPTIC FILTER

Sangeetha Marikkannan

Karpaga Vinayaga College of Engineering and Technology, Chennai

sang_gok @yahoo.com

***ABSTRACT.***
*A new evolutionary algorithm for scheduling and allocation algorithm is developed for an elliptic filter. The elliptic filter is scheduled and allocated in the proposed work which is then compared with the different scheduling algorithms like As Soon As Possible algorithm, As Late As Possible algorithm, Mobility Based Shift algorithm, FDLS, FDS and MOGS. In this paper execution time and resource utilization is calculated using different scheduling algorithm for an Elliptic Filter and reported that proposed Scheduling and Allocation increases the speed of operation by reducing the control step. The proposed work to analyse the magnitude, phase and noise responses for different scheduling algorithm in an elliptic filter.*

## 1. INTRODUCTION

In this process, the scheduling and allocation algorithm is designed with an objective to minimize the control steps which in turn reduce the cost function. Elliptic filter is a signal processing filter with equalized ripple behavior in both the pass band and stop band. The process tasks are scheduling and allocation[1]. The first step in the scheduling and allocation algorithm design, which is the case of transforming elliptic filter program into structure, includes operation scheduling and resource allocation. The scheduling and allocation algorithm are closely interrelated. In order to have an optimal design, both task should be performed simultaneously. However, due to time complexity, many systems perform them separately or introduce iteration loops between the two subtasks. Scheduling involves assigning the operation to so- called control steps. A control step is the fundamental sequencing unit in the synchronous system; it corresponds to a clock cycle.

Allocation involves assigning the operation and values to resources i.e., providing storage, functional units and communication paths are specifying their usage. Therefore, allocation is usually further divided into three subtasks: variable binding, operation assignment and data transfer binding. Variable binding refers to the allocation of register to data, i.e., values that are generation one control step and used in another must be assigned to registers. Few systems have a one-to one correspondence between variables and registers, while other allow register sharing for those variables, which have disjoint life times. Operation assignment binds operation to functional units

(e.g., an adder or an ALU). Data transfer bindings represent the allocation of connections (e.g., buses, multiplexer) between hardware components i.e., registers and functional units to create the necessary information paths as required by the specification and the schedule.

There is a variety of scheduling algorithms that differ in the way of searching for the best solution. Mostly they optimize only the number of functional units. In our evaluation process, it turned out that scheduling algorithm obtained the best results in terms of utilization of functional units and computation time [2]. Therefore it is reasonable to use the principles of the evolutionary algorithm to find some of the optimum solutions.

However, there are different approaches to sewing the same problem, but it is not important how close the algorithm comes to the optimum solutions: what matters is how those schedules are allocated in the final design. Therefore, since the subtasks of scheduling and allocation are heavily interrelated, the algorithm cannot be judged in terms of optimization until the final result of the allocation subtask is known. So, when a new scheduling algorithm is created the allocation criteria has to be taken in to account [3].

The main steps involved in scheduling and allocation algorithm for an elliptic filter.

Description of the behaviour of the system.

Translation of the description in to a graph (e.g., CDFG)[5].

Operation scheduling (each operation in the graph is assigned to a control step).

Allocation of the resources for the digital system (here the resources can be functional units assigned to execute operation derived from the graph CDFG).

Portioning the system behaviour in to the hardware and software module for the scheduled CDFG to estimate buffer size and delay.

Usually, allocation, scheduling and assignment are widely recognized as mandatory backbone tasks in high-level synthesis.

## 2. PREVIOUS APPROACH

Techniques for combined scheduling and check point insertion in high-level synthesis of digital systems. More generalized CDFGs are needed to be designed. Ravi kumar (1998) present an adaptive version of the well-known simulated annealing algorithm and described application to a combinatorial optimization problem arising in the high level synthesis of digital systems. It takes 29 registers for the 5 functional registers[4]. Scheduling method for reducing the number of scan register for a cyclic structure. In order to estimate the number of scan register during scheduling, and they proposed a provisional binding of operational units and showed a force-directed scheduling algorithm with the provisional binding [2] cluster based register binding is performed that binds each carrier of DFG to a register. A set of resources consisting of functional units and registers is assigned to each cluster, and instead of binding the resources to and sharing them among individual operations or carriers, the set of resources is bound to and shared among the clusters. Such as approach, help to reduce the average active cycles of those clocked elements [3]. Multi objective genetic scheduling(MCGS) algorithm which shows less cost function than other scheduling algorithm for an elliptic wave filter. The proposed scheduling and allocation algorithm shows less execution time and cost function [1].

## 3. NODES ASSIGNMENT

In general the nodes in a CDFG can be classified as one of the following types.

Operational nodes: These are responsible for arithmetic logical or relational operations.

Call nodes: This node denotes calls to sub program modules.

Control nodes: This node is responsible for applications like conditional and loop constructs.

Storage nodes: These nodes represent assignment applications associated with variables and signals.

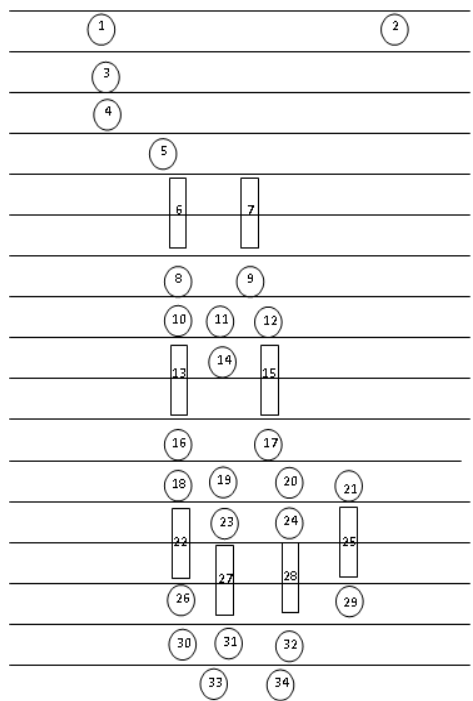

**Fig.1.**Operational nodes for Elliptic wave filter benchmark

## 4. CDFG GENERATION

The control data flow graph is a directed a cyclic graph in which a node can be either an operation node or a control node (representing a branch, loop, etc.,).the directed edges in a CDFG represent the transfer of a value or control from one node to another. An edge can be condition, while implementing a if/care statements or loop constructs[4].The Table 1 shows the estimated functional units and registers for Hardware Oriented Approach.

**Table.1.** Hardware oriented approach

|  |  |  | Functional unit | Register |
|---|---|---|---|---|
| Binary partitioning | Proposed work | merging | 1 | 4 |
| Source level partitioning | Papa&silc(2000) | ASAP | 8 | 11 |
|  | Papa&silc(2000) | ALAP | 9 | 13 |
|  | Papa&silc(2000) | FDS | 6 | 11 |
|  | Papa&silc(2000) | LS | 6 | 11 |
|  | Papa&silc(2000) | FDLS | 6 | 11 |
|  | Papa&silc(2000) | MOGS | 6 | 11 |
|  | Proposed work | SAA | 5 | 11 |
|  | Deniz Dal &NazaninMansouril(2008) | - | 5 | 12 |

## 5. HARDWARE COMPONENTS

A very essential hardware component is the functional unit (FU). This is a combinatorial or sequential logic circuit that realizes Boolean functions, such as an adder, a multiplexer or an arithmetic unit( ALU). Another essential component inherent to synchronous logic is the register, which makes it possible to store data in the circuit. Both FUs and registers operate on words, which means that each input or output is actually realized by a number of signals carrying a bit. A register is a simplest form of a memory element, separate registers are not very efficient, as individual wires have to be connected to each register. Registers are then combined to form register files. The simplest way to make connections between hardware components is by using wires. The hardware can be used much efficiently; it is possible to change the connection between the components during a computation. One way to realize this is to use multiplexing. Even more efficient hardware design is possible if buses and tri-state drivers are used.

## 6. SCHEDULING

Scheduling algorithm can be constructive or transformational (based on their approach). Transformational algorithms start with some schedule (typically parallel or maximally serial) and separately apply transformations in an attempt to bring the schedule closer to the design requirements[3]. The transformations allow operations to be parallelized or serialized whilst ensuring that dependency constraints between operations are not violated. In contrast, constructive algorithm builds up a schedule from scratch by incrementally adding operations. A simplest e.g., of the constructive approach scheduling is As Soon As Possible (ASAP) scheduling.

## 7. SCHEDULING AND ALLOCATION ALGORITHM

The data flow graph obtained from the input description to scheduled using As Soon As Possible scheduling and As Late As Possible scheduling. ASAP scheduling computes the earliest time at which an operation can be scheduled and ALAP can also be computed by adapting the longest path algorithm to work from the output backwards . Combining the information obtained in both ways of scheduling algorithm give rise to more powerful heuristics called mobility based scheduling. (according to the available functional units).

 The scheduling algorithm proposed take care of resource constrained synthesis and find a scheduling and assignment, such that, the total computation is completed in minimal time(resource constrained synthesis). The proposed scheduling reduces the critical path of the data flow graph.

  The root nodes are calculated from the graphical description and the critical path is determined. The algorithm merges the node that has data dependency, which is of same type and has minimum two extend an input that is decomposed in to parallel form.

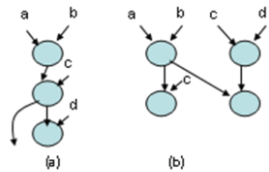

**Fig.2.** (a) Most serial form  (b) Most parallel form

The condition is that, last node should have single output edge, if predecessors have one output edge than the both nodes are merged and formed in to single node. If the node has more than one output edge the node should not be disturbed and a cut in the path is set and the current node is moved to previous cycle where it meets the hardware constraint problem are shown in Fig.2. If the problem satisfies the condition, a node is inserted in the previous cycle else it chooses the critical path. If the critical path is cut, the control step of the CDFG is reduced, which leads to a reduction in clock cycle of the entire system without any change in the hardware constraint.

 A cut in the critical path, i.e., between node 3and node 4 converts, the most serial is converted in to most parallel form, and leads to a reduction in a single control step without affecting the hardware constraints of 3 adder and 2 multiplier. Hardware is allocated according to data dependency of the nodes are shown in Fig.3.

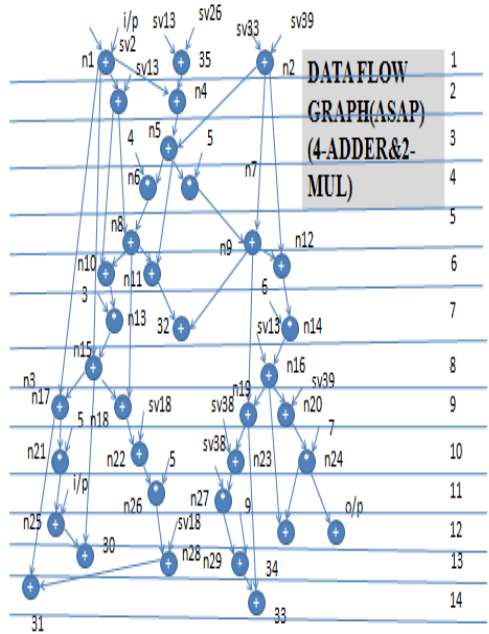

**Fig.3.** Scheduling and Allocation Algorithm for Elliptic Wave Filter

## 8. EXPERIMENTAL RESULTS

The proposed scheduling algorithm cost less and also has reduced control steps when compared to other scheduling algorithms and utilizes minimum number of hardware resources. A cut established between hardware and software partitioned nodes determines the edge cut and buffer size is determined from the life time of the edge. The proposed scheduling and allocation algorithm proves to achieve better solution for two way portioning. Table 2 shows the control step and Execution time. Table 3 represents the software Oriented Approach using ARM processor.

**Table.2.** Software oriented approach using LABVIEW

| Algorithm | Number of Functional Unit | | Number of Control Step | Number of Execution Time (ms) |
|---|---|---|---|---|
| | + | * | | |
| ASAP | 3 | 2 | 16 | 34.24 |
| ALAP | 3 | 2 | 16 | 38.88 |
| MBS | 3 | 2 | 16 | 39.52 |
| SAA | 4 | 2 | 13 | 22.56 |

**Table.3.** Software oriented approach using ARM processor

| Algorithm | No. of Ripple | No. of cluster | Ripple execution time (0-1000) ms | Cluster Execution time (0-1000) ms |
|---|---|---|---|---|
| ASAP | 6 | 4 | 32 | 98 |
| ALAP | 8 | 5 | 13 | 45 |
| MBS | 2 | 8 | 10 | 19 |
| SAA | 2 | 1 | 209 | 568 |

## 9. ASAP SCHEDULING ALGORITHM

In this Front panel of ASAP scheduling algorithm using ARM processor the analog inputs are connected and more ripple and cluster are analysed in between (0-1000) execution time. The number of ripple in ASAP is 6 and ripple execution time is 32ms. The number of cluster in between (0-1000) is 4 and cluster execution time is 98ms are shown in fig.4.

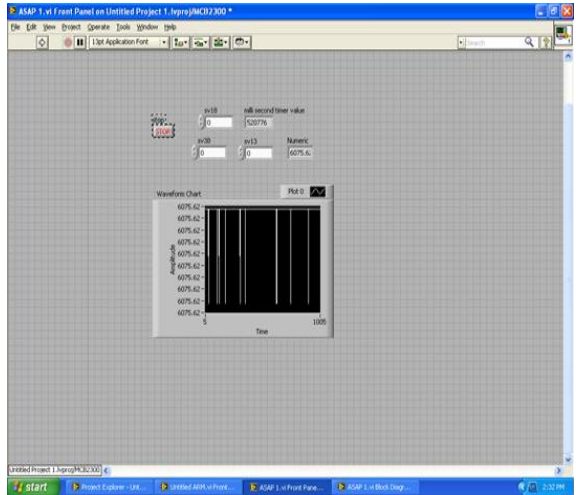
**Fig.4.** Response of ASAP using ARM Processor

In this Front panel of ASAP scheduling algorithm the first figure shows the magnitude response are analysed using LABVIEW in which the ripple frequency is 10MHZ. The second figure shows the phase response of ASAP in which the ripple frequency is 11MHZ and third figure represents the noise response of ASAP are shown in fig.5..

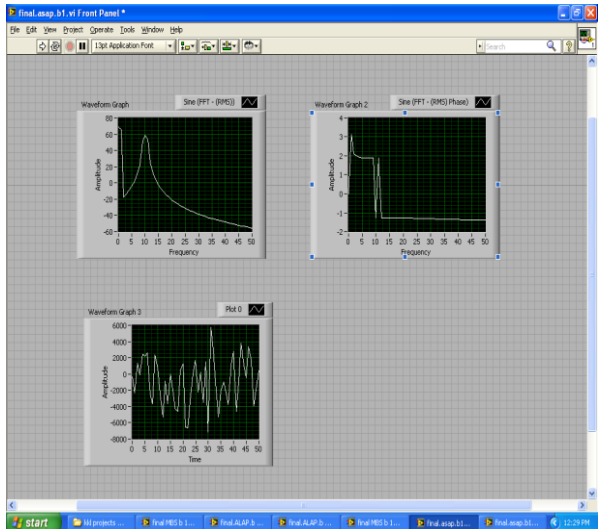
**Fig.5.** Response of of ASAP using LABVIEW

## 10. ALAP SCHEDULING ALGORITHM

In this Front panel of ALAP scheduling algorithm using ARM processor the analog inputs are connected and more ripple and cluster are analysed in between (0-1000) execution time. The number of ripple in ALAP is 8 and ripple execution time is 24ms. The number of cluster in between (0-1000) is 5 and cluster execution time is 45ms are shown in fig.6.

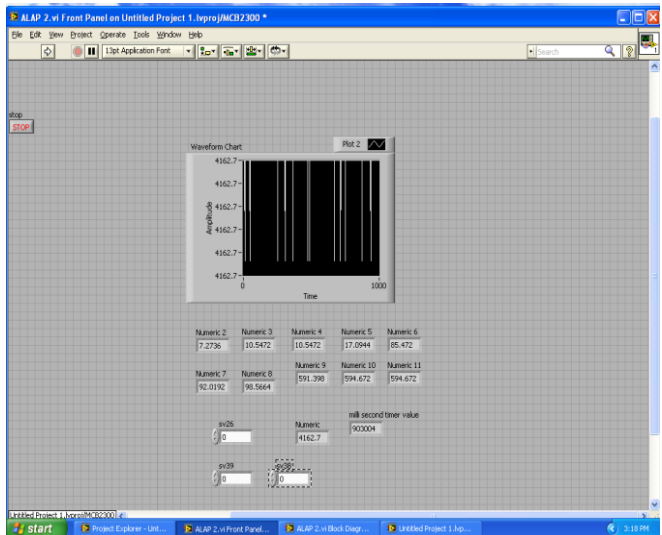
**Fig.6.** Response of ALAP using ARM processor

In this Front panel of ASAP scheduling algorithm the first figure shows the magnitude response are analysed using LABVIEW in which the ripple frequency is 10MHZ. The second figure shows the phase response of ASAP in which the ripple frequency is 11MHZ and third figure represents the noise response of ASAP are shown in fig.7.

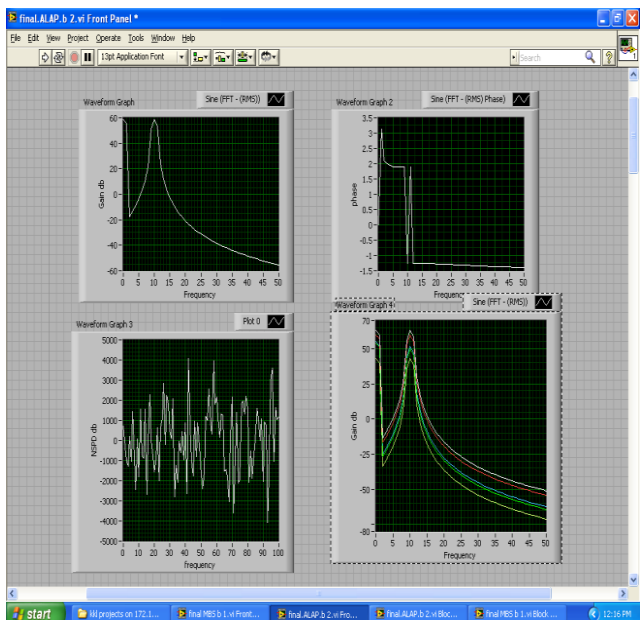
**Fig.7** Responses of ASAP using LABVIEW

The response of Mobility based Scheduling is represented in the Figure.8 which shows cluster and ripple in the pass band. It shows 2 ripple and 8 cluster in 1005 ms. In Figure 9 shows the implementation of Mobility based Scheduling using ARM processor (MCB2300). Figure 10 shows the Scheduling and Allocation Algorithm using Labview and Figure 11 shows the implementation of Scheduling and Allocation Algorithm using ARM Processor(MCB2300). For Scheduling and Allocation Algorithm there exists 2 ripple and 1 cluster. The ripple exists in 209 ms which is represented in Figure 11.

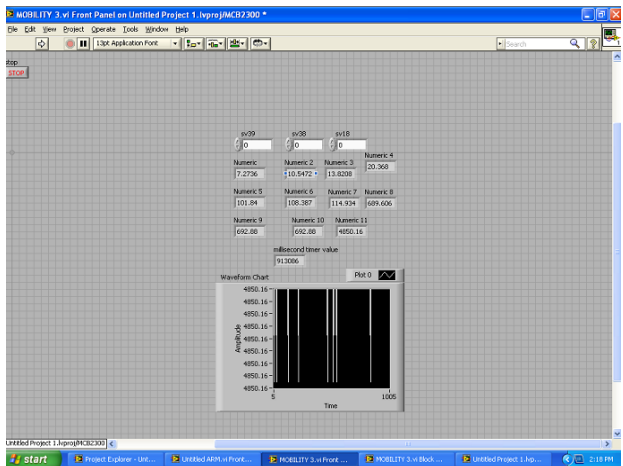
Fig.8. Response of MBS using Labview

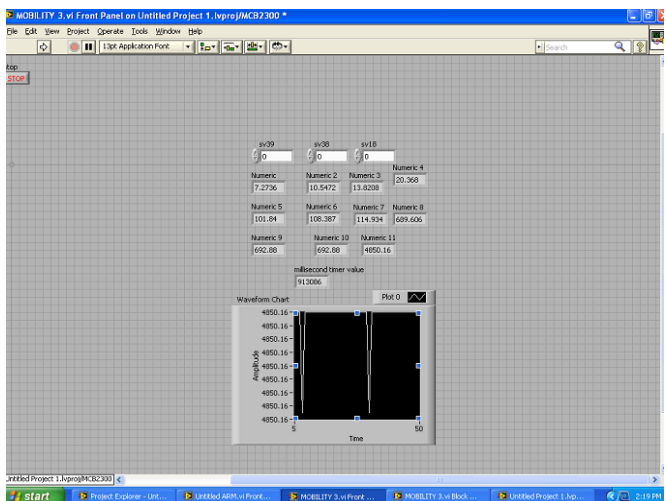
Fig.9. Response of MBS after implementation in ARM Processor

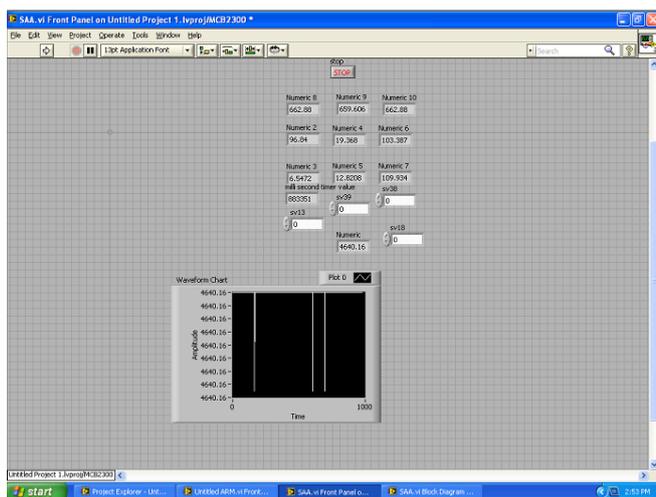
Fig.10. Response of Scheduling Allocation Algorithm using Labview

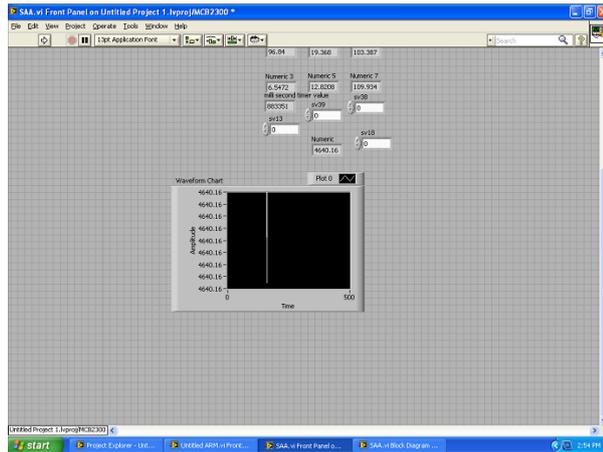

Fig.11. Response of Scheduling and Allocation Algorithm using ARM Processor

## 11. CONCLUSIONS

The scheduling and allocation algorithm is designed by converting most serial form in to most parallel form and placing a cut in the serial path which leads to a decrease in the critical path length without any change in functionality. The scheduling and allocation algorithm proposed reduces the execution time and cost function by reducing control step (one step). An effort is not made to reduce the critical path length in the earlier reported works. However, the use of scheduling and allocation algorithm shortens the control step to 16 without modifying the functionality. Four partitioning methods are performed for scheduled control data flow graph. In the first method, the partitioning is done such that, the operation that takes more number of cycles is placed in hardware units. Other methods uses clique partitioning to minimize the number of resources used. Buffer size and system delay for hardware/ software partitioning is also calculated to obtain communication cost.

**References**

1. Papa, G.; Silc, J, Multi-objective genetic scheduling algorithm with respect to allocation in high-level synthesis, Proceedings of the 26th Euromicro Conference, Volume 1, 5-7 Sept 2000 pp: 339 – 346 (2002)
2. Gregor Papa*and Jurij Silc, "Automated large-scale integrated synthesis using allocation-based scheduling algorithm", IEEE 2 Feb. (2002)
3. Myoung-Keun You and Gi-Yong Song, "Implementation of a C-to-SystemC Synthesizer Prototype", School of Electrical and Computer EngineeringChungbuk National University, Cheongju Chungbuk, 361-763, Korea (2007)
4. John B. Hughes, Kenneth W. Moulding, Judith Richardson, John Bennett, William Redman-White, Mark Braceyand Randeep Singh Soin, "Automated Design of Switched-Current Filter", IEEE Journal of Solid State Circuits, VOL. 31, NO. 7, JULY (1996)
5. John Hansen and Montek Singh, A Fast Branch-and-Bound Approach to High-Level Synthesis of Asynchronous Systems, IEEE Symposium on Asynchronous Circuits and Systems,(2010).

**Sangeetha Marikkannan** born in 1975 in Tamilnadu, India, received his B.E. degree from the Bharathidasan University, Trichy, in

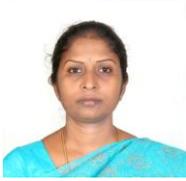

1996 and M.E. degree from the University of Madras in 1999. She is currently with Karpaga Vinayaga College of Engineering and Technlogy in the Department of Electronics and Communication Engineering and Ph.D. degree with the Anna University, Chennai, India. She is a member of the IEEE and ISTE. Her research interests are Hardware/Software Codesign and High Level Synthesis.